\documentclass{article}
\usepackage{spconf,amsmath,graphicx, color, subfigure}

\title{\normalsize COMBINING NONPARAMETRIC SPATIAL CONTEXT PRIORS WITH NONPARAMETRIC SHAPE PRIORS FOR DENDRITIC SPINE SEGMENTATION IN 2-PHOTON MICROSCOPY IMAGES}
\name{\normalsize Ertunc Erdil$^{1, 2}$ \quad A. Ozgur Argunsah$^{3}$ \quad Tolga Tasdizen$^{4}$ \quad Devrim Unay$^{5}$ \quad Mujdat Cetin$^{1, 6}$ \thanks{This work has been supported by the Scientific and Technological Research Council of Turkey (TUBITAK) under Grant 113E603.}}
 \address{\normalsize $^{1}$ Faculty of Engineering and Natural Sciences, Sabanci University, Istanbul, Turkey \\
 \normalsize $^{2}$ ARM Ltd., 1 Summerpool Road, Loughborough, Leicester, UK \\
\normalsize $^{3}$ Brain Research Institute, University of Zurich, Zurich, Switzerland \\
\normalsize $^{4}$ Department of Electrical and Computer Engineering, University of Utah, Salt Lake City, UT, USA\\
\normalsize $^{5}$ Department of Biomedical Engineering, Izmir University of Economics, Izmir, Turkey\\
\normalsize$^{6}$ Department of Electrical and Computer Engineering, University of Rochester, Rochester, NY, USA\\
\normalsize \{ertuncerdil, mcetin\}@sabanciuniv.edu \quad argunsah@hifo.uzh.ch \quad tolga@sci.utah.edu \quad devrim.unay@ieu.edu.tr\\
}
\begin{document}
%\ninept
%
\maketitle
\begin{abstract}
%Combination of nonparametric shape priors with a data term in a Bayesian framework has led to successful segmentation results in scenarios involving challenging image data that provide limited information. 
Data driven segmentation is an important initial step of shape prior-based segmentation methods since it is assumed that the data term brings a curve to a plausible level so that shape and data terms can then work together to produce better segmentations. When purely data driven segmentation produces poor results, the final segmentation is generally affected adversely. One challenge faced by many existing data terms is due to the fact that they consider only pixel intensities to decide whether to assign a pixel to the foreground or to the background region. When the distributions of the foreground and background pixel intensities have significant overlap, such data terms become ineffective, as they produce uncertain results for many pixels in a test image. In such cases, using prior information about the spatial context of the object to be segmented together with the data term can bring a curve to a plausible stage, which would then serve as a good initial point to launch shape-based segmentation. In this paper, we propose a new segmentation approach that combines nonparametric context priors with a learned-intensity-based data term and nonparametric shape priors. We perform experiments for dendritic spine segmentation in both 2$D$ and 3$D$ 2-photon microscopy images. The experimental results demonstrate that using spatial context priors leads to significant improvements.
\end{abstract}

\begin{keywords}
Nonparametric shape priors, spatial context priors, spine segmentation, 2-photon microscopy.
\end{keywords}
\section{Introduction}
\label{sec:intro}
Segmentation of images involving limited and low quality data is a challenging problem and requires prior information about the shapes of the objects to be segmented for accurate results \cite{erdil2017nonparametric, cremers2006kernel, kim2007nonparametric}. Earlier work involved the use of curve-length penalties, essentially providing simple prior information about shape regularity \cite{kass1988snakes}. Later, by using linear analysis tools such as principal component analysis (PCA), more informative shape priors learned from training samples have been incorporated into the segmentation process \cite{tsai2003shape}. However, these methods can only handle Gaussian-like, unimodal, shape prior densities. In order to handle multimodal shape densities, methods that exploit nonparametric shape priors have been proposed \cite{erdil2017nonparametric, cremers2006kernel, kim2007nonparametric, erdil2016mcmc}. The major deficiency of these techniques is that they often combine such priors with simplistic data terms. A common underlying assumption of such data terms is that the foreground and the background regions in the image are homogeneous, e.g., intensities are piecewise constant or piecewise smooth \cite{souganli2014combining}. Learning-based intensity distributions have been integrated as data terms with nonparametric shape priors to handle more complicated intensity distributions \cite{souganli2014combining}.

Integration of nonparametric shape priors with learned-intensity-based data terms has led to improved segmentation results when the foreground and background intensities are not homogeneous \cite{souganli2014combining}. Data driven segmentation is an important initial step of segmentation methods that exploit shape priors. In learned-intensity-based data terms, intensity distributions of the foreground and background regions are learned from a training set, possibly in a nonparametric fashion. In challenging segmentation tasks, the foreground and background intensity distributions have significant overlap, causing the segmentation algorithm to produce uncertain region assignments for many pixels, leading to poor segmentation results.

\begin{figure}[ht]
\centering
\subfigure[Input image.\label{fig:toy_example_test}]{\includegraphics[scale=0.2]{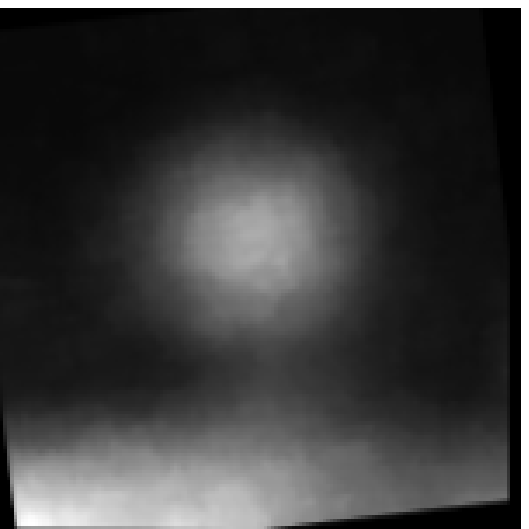}}
\subfigure[Ground truth.\label{fig:toy_example_gt}]{\includegraphics[scale=0.2]{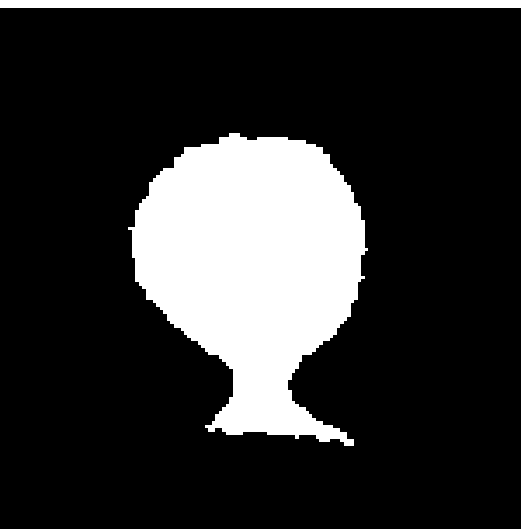}}
\subfigure[Initialization.\label{fig:toy_example_initialization}]{\includegraphics[scale=0.2]{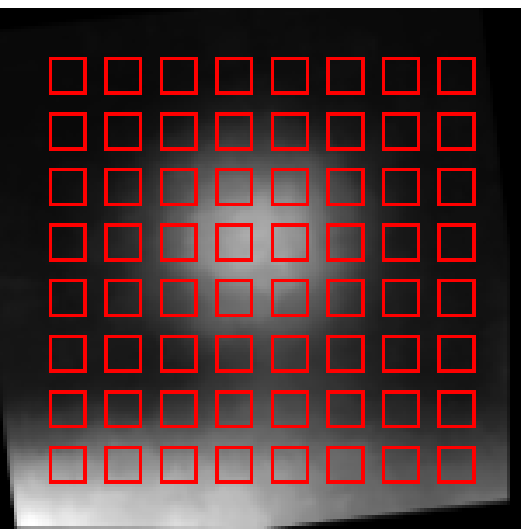}}
\subfigure[Purely data-driven\label{fig:toy_example_intensity}]{\includegraphics[scale=0.2]{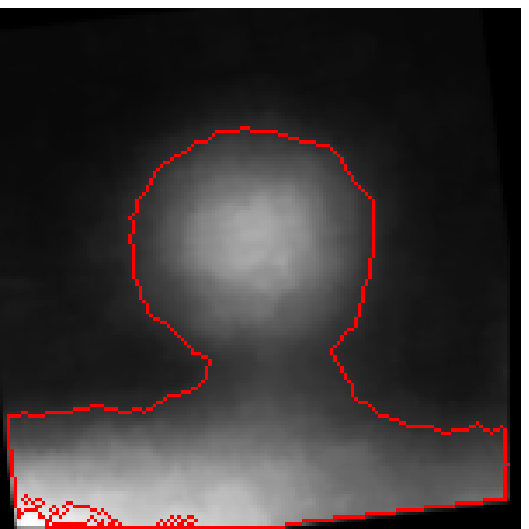}}
\subfigure[Context priors\label{fig:toy_example_intensity_context}]{\includegraphics[scale=0.2]{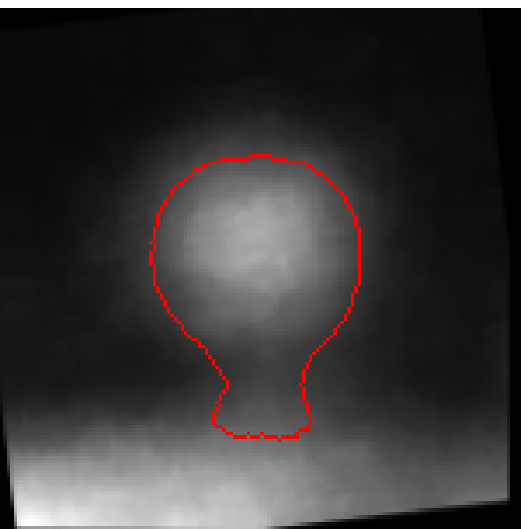}}
\subfigure[Shape priors\label{fig:toy_example_intensity_shape}]{\includegraphics[scale=0.2]{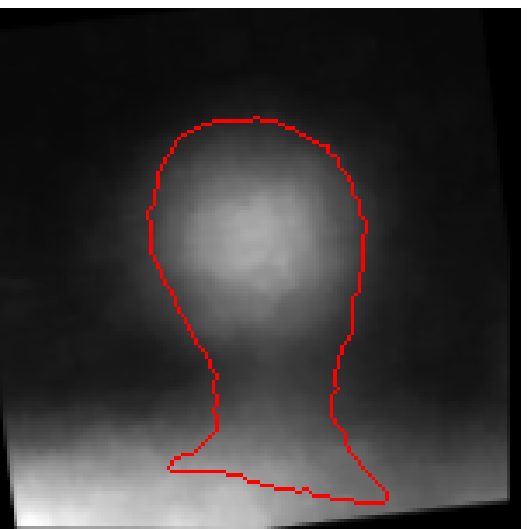}}
\subfigure[Context and shape priors\label{fig:toy_example_intensity_context_shape}]{\includegraphics[scale=0.2]{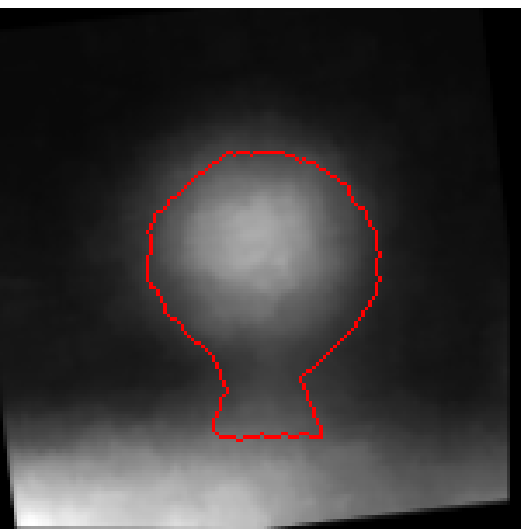}}
\subfigure[Probability density functions.\label{fig:toy_example_pdf}]{\includegraphics[scale=0.13]{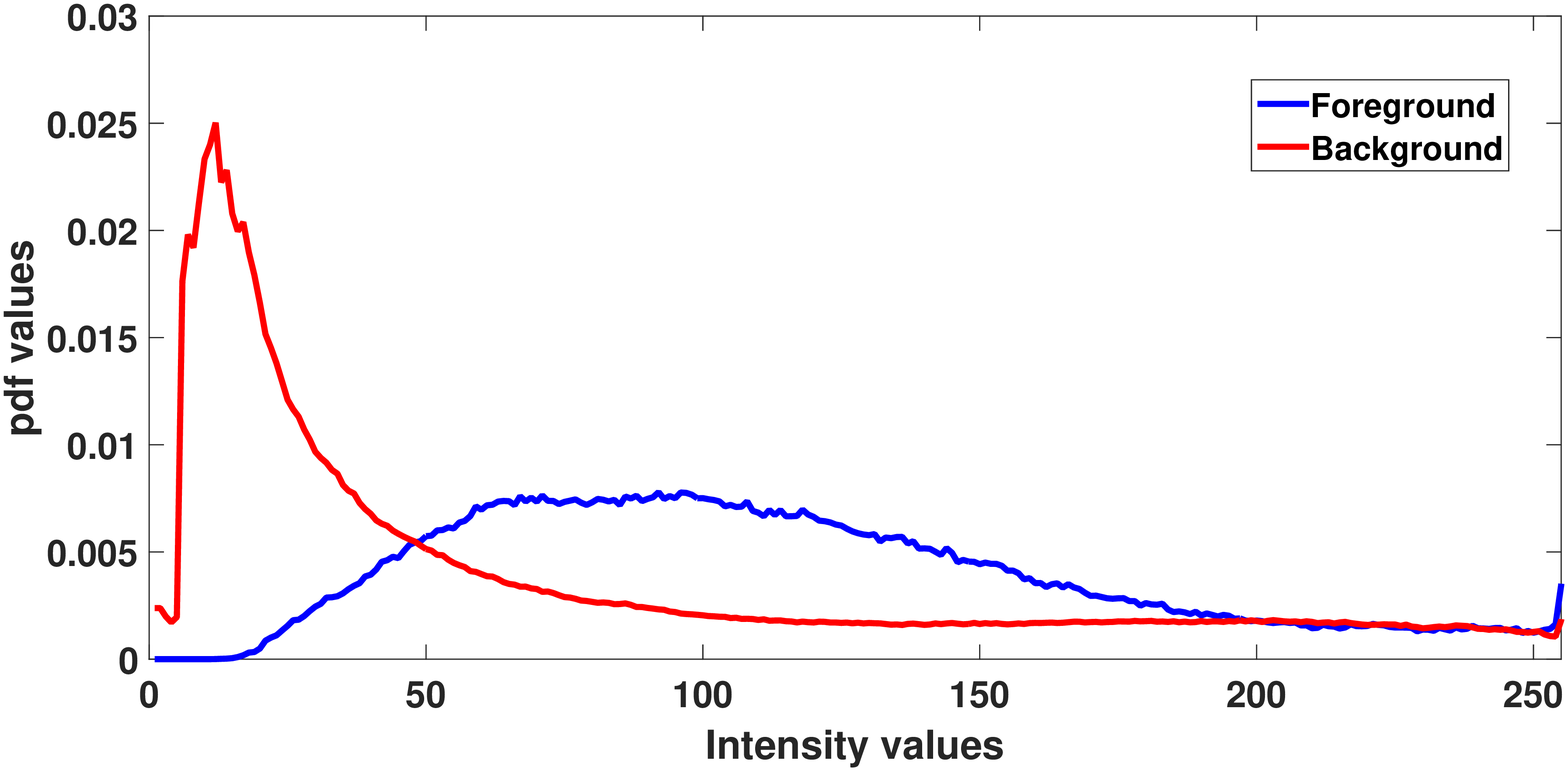}}
\caption{Motivation of the proposed approach on an example.}
\end{figure}

Let us consider the 2-photon microscopy image of a dendritic spine in Figure \ref{fig:toy_example_test} as an illustrative example. Dendritic spines are small protrusions from a neuron's dendrite, and are of interest in neuroscience research as their density and morphology are related to several functions including learning and memory. For analysis of their morphology, their segmentation is crucial. The probability density functions (pdfs) learned from a training set for the foreground and background intensities are shown in Figure \ref{fig:toy_example_pdf}. Note that bright intensities might appear with almost equal probabilities in both the foreground and background regions according to the pdfs. Since data-driven segmentation is performed based on these intensity distributions, bright pixels from the foreground region may easily be assigned to the background region and vice versa. Segmentation of the dendritic spine image using a learned-intensity-based data term produces the segmentation result shown in Figure \ref{fig:toy_example_intensity}. Note that some bright pixels are assigned to foreground region, while they should be in the background. Spatial context priors incorporate higher level of information about the pixel locations of the objects to be segmented. Therefore, by using spatial context priors (as described later in this paper) together with learned-intensity-based data term, we exploit information about the location of pixels in the image together with the intensity pdfs. This produces the segmentation result shown in Figure \ref{fig:toy_example_intensity_context}. The result obtained by using context priors together with learned-intensity-based data term is closer to the ground truth (see Figure \ref{fig:toy_example_gt}) than the one obtained by purely data-driven segmentation. Therefore, the segmentation in Figure \ref{fig:toy_example_intensity_context} would be a better initialization for shape-based segmentation. Giving the boundary in Figure \ref{fig:toy_example_intensity} as initialization to an algorithm using a learned-intensity-based data term and shape priors produces the result in Figure \ref{fig:toy_example_intensity_shape} which is not quite similar to the ground truth. On the other hand. the approach developed in this paper, exploiting both spatial context and shape priors, initialized by the boundary in Figure \ref{fig:toy_example_intensity_context} produces a closer segmentation result to the ground truth as shown in Figure \ref{fig:toy_example_intensity_context_shape}. 

In this paper, we propose a new segmentation approach that uses spatial context priors together with shape priors and a learned-intensity-based data term in a Bayesian framework. We learn all the densities involved in a nonparametric fashion from a training set. To the best of our knowledge such spatial context priors have not been used together with learned-intensity-based data terms and nonparametric shape priors for segmentation in the literature. We perform experiments for dendritic spine segmentation on both 2$D$ and 3$D$ 2-photon microscopy images and demonstrate the improvements provided by the proposed approach.

This approach would be of potential use in challenging biomedical and biological image segmentation problems (e.g. brain tumor segmentation, prostate segmentation, etc.) in which one can collect statistical information regarding the shapes, intensities, and locations of the particular structures of interest, from training data. Our framework provides a principled probabilistic formulation and an associated algorithm to exploit such information.

\section{Proposed method}
Let us assume that we have a training set of $n$ aligned intensity images $\mathcal{Y} = \{y_1, \cdots, y_n\}$ and their corresponding manual segmentations, $\mathcal{C} = \{c_1, \cdots, c_n\}$. Given $\mathcal{Y}$ and $\mathcal{C}$, we can construct multisets\footnote{We use multiset to indicate a set that contains repeated values.}, $I_{fg}$ and $I_{bg}$, that store all intensity values for pixels in the training set that are located in the foreground and background regions, respectively. Similarly, we can construct multisets, $L_{fg}$ and $L_{bg}$, that store the locations of all pixels in the foreground and background regions, respectively. We define the posterior probability density function of segmenting curve $c$ given observed image $y$ and spatial context of the object to be segmented, $l$ as
\begin{equation}
p(c|y, l) \propto p(y, l | c) p(c) \propto p(y|c) p(l|c) p(c).
\label{eq:posterior}
\end{equation}
where $y$ and $l$ contain the pixel intensities and the associated locations. Hence our formulation exploits information, gathered from from training data, not only about intensities of pixels in different regions but also about locations of pixels belonging to different regions in a principled probabilistic framework.

By taking the negative logarithm of Equation \eqref{eq:posterior}, we can define the following energy function to be minimized for segmentation:
\begin{equation}
E(c) = -\log p(y|c) - \log p(l|c) - \log p(c).
\label{eq:energy}
\end{equation}

In this paper, we use level sets as shape representation. Level set representation is essentially a mapping, $\phi$ from the binary space to the real space. In the literature, it has been found more convenient to work with level sets to handle topological shape changes \cite{fedkiw2002level} and computing gradients \cite{kim2007nonparametric}. In the level set representation we use, values less than zero indicate foreground region whereas values greater than zero indicate background region. Using level set representation, the energy function in Equation \eqref{eq:energy} becomes
\begin{equation}
E(\phi(c)) = -\log p(y|\phi(c)) - \log p(l|\phi(c)) - \log p(\phi(c)).
\label{eq:energy_level_set}
\end{equation}

We define $-\log p(y|\phi(c))$ as proposed in \cite{souganli2014combining}:
\begin{equation}
\scriptsize
-\log p(y|\phi(c)) = -\int_{\phi(c) < 0} p_{fg}(y(x))\mathrm{d}x - \int_{\phi(c) > 0} p_{bg}(y(x))\mathrm{d}x
\end{equation}
where
\begin{equation}
\scriptsize
p_{fg}(y(x)) = \frac{1}{|I_{fg}|} \sum_{i = 1}^{|I_{fg}|} \mathcal{N}(y(x); I_{fg}(i), \sigma)
\label{eq:p_fg}
\end{equation}
and
\begin{equation}
\scriptsize
p_{bg}(y(x)) = \frac{1}{|I_{bg}|} \sum_{i = 1}^{|I_{bg}|} \mathcal{N}(y(x); I_{bg}(i), \sigma).
\label{eq:p_bg}
\end{equation}
In Equations \eqref{eq:p_fg} and \eqref{eq:p_bg}, $\mathcal{N}(.; \mu, \sigma)$ indicates a Gaussian density with mean $\mu$ and standard deviation $\sigma$ and $I_{fg}(i)$ ($I_{bg}(i)$) indicates $i^{th}$ element of $I_{fg}$ ($I_{bg}$). Hence \eqref{eq:p_fg} and \eqref{eq:p_bg} provide nonparametric pdf estimates of intensities in the foreground and background regions.

Similarly, we define $-\log p(l|\phi(c))$ as:
\begin{equation}
\scriptsize
-\log p(l|\phi(c)) = -\int_{\phi(c) < 0} q_{fg}(x)\mathrm{d}x - \int_{\phi(c) > 0} q_{bg}(x)\mathrm{d}x
\end{equation}
where
\begin{equation}
%\scriptsize
q_{fg}(x) = \frac{1}{|L_{fg}|} \sum_{i = 1}^{|L_{fg}|} \mathcal{N}(x; L_{fg}(i), \sigma)
\label{eq:q_fg}
\end{equation}
and
\begin{equation}
%\scriptsize
q_{bg}(x) = \frac{1}{|L_{bg}|} \sum_{i = 1}^{|L_{bg}|} \mathcal{N}(x; L_{bg}(i), \sigma)
\label{eq:q_bg}
\end{equation}

Finally, we define $p(\phi(c))$, the shape prior density, as $p(\phi(c)) = \frac{1}{n} \sum_{i = 1}^n \mathcal{N}(\phi(c); \phi(c_i), \sigma I)$ where $\mathcal{N}(.; \mu, \sigma I)$ indicates a Gaussian density with mean vector $\mu$ and covariance matrix $\sigma I$, and $I$ is the identity matrix.

The segmentation problem turns into the problem of finding a boundary $c$ that minimizes the energy functional in Equation \eqref{eq:energy}. To achieve this, we minimize Equation \eqref{eq:energy_level_set} using gradient descent which requires computing the partial derivative of $E(\phi(c))$ with respect to $\phi(c)$. This is equivalent to computing partial derivatives of each component involved which are written as
\begin{equation}
\frac{- \partial \log p(y(x) | \phi(c))}{\partial \phi(c)} = \log \frac{p_{bg}(y(x))}{p_{fg}(y(x))},
\label{eq:partial_data}
\end{equation}

\begin{equation}
\frac{- \partial \log p(l(x) | \phi(c))}{\partial \phi(c)} = \log \frac{q_{bg}(x)}{q_{fg}(x)},
\label{eq:partial_context}
\end{equation}
and
\begin{equation}
%\begin{split}
\scriptsize
\frac{\partial -\log p(\phi(c))}{\partial \phi(c)} = \frac{1}{p(\phi(c))} \frac{1}{\sigma^2} \frac{1}{n}\sum_{i = 1}^n \mathcal{N}(\phi(c); \phi(c_i), \sigma I) \times (\phi(c_i) - \phi(c))
%\end{split}
\label{eq:partial_shape}
\end{equation}

Given partial derivatives of each term, we perform curve evolution in the gradient direction of these terms to find the desired segmentation. The algorithmic steps of the proposed segmentation approach are: (i) Perform curve evolution until convergence using the learned-intensity-based data term and the context priors terms, i.e., \eqref{eq:partial_data} and \eqref{eq:partial_context}; (ii) Take the curve found in (i) as initialization and perform curve evolution until convergence using the learned-intensity-based data term together with the spatial context and shape priors terms, i.e., \eqref{eq:partial_data}, \eqref{eq:partial_context}, \eqref{eq:partial_shape}.
\begin{table}[ht]
\centering
\caption{Average Dice score results on both 2$D$ and 3$D$ dendritic spine data sets}
\label{tab:dice_results}
\resizebox{.5\textwidth}{!}{
\begin{tabular}{|c|c|p{5cm}|c|c|c|}
\hline                                                                                  & \textbf{\begin{tabular}[c]{@{}l@{}}Proposed\\ method\end{tabular}} & \textbf{\begin{tabular}[c]{@{}l@{}}Context priors with\\  learned-intensity-based data term\end{tabular}} & \textbf{\cite{chan2001active}} & \textbf{\cite{kim2007nonparametric}} & \textbf{\cite{souganli2014combining}} \\ \hline
\textbf{\begin{tabular}[c]{@{}l@{}}3$D$ Dendritic \\spine data set\end{tabular}} & 0.7022                   & 0.6597                                                                        & 0.0989                                         & 0.4319                                               & 0.5857                                                \\ \hline
\textbf{\begin{tabular}[c]{@{}l@{}}2$D$ Dendritic \\spine data set\end{tabular}}  & 0.8590                   & 0.8341                                                                        & 0.5188                                         & 0.5759                                               & 0.7347                                                \\ \hline
\end{tabular}
}
\end{table}
\begin{figure}[ht]
\centering
\includegraphics[scale=0.2]{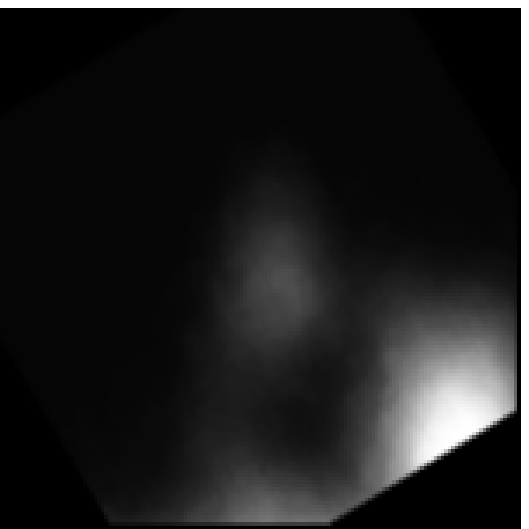}
\includegraphics[scale=0.2]{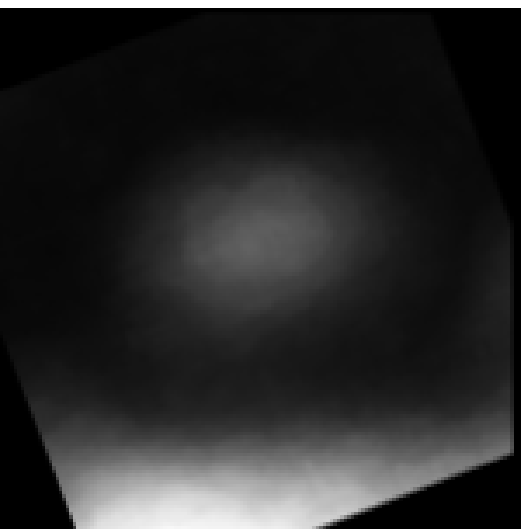}
\includegraphics[scale=0.2]{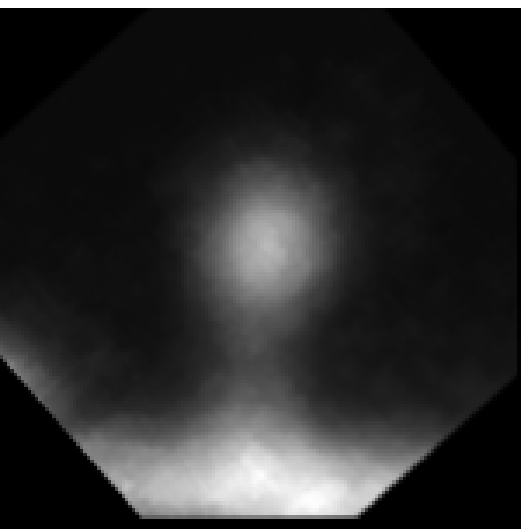}
\includegraphics[scale=0.2]{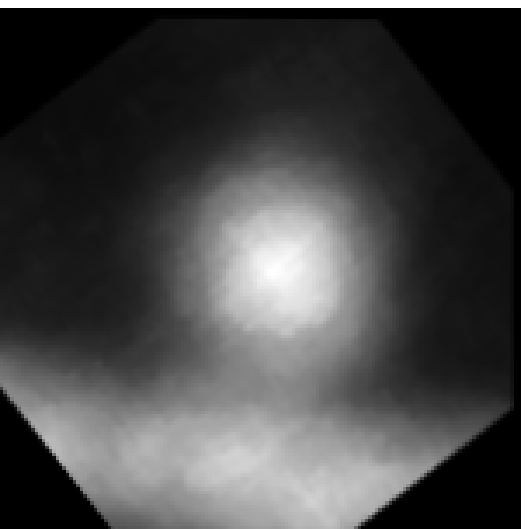}
\includegraphics[scale=0.2]{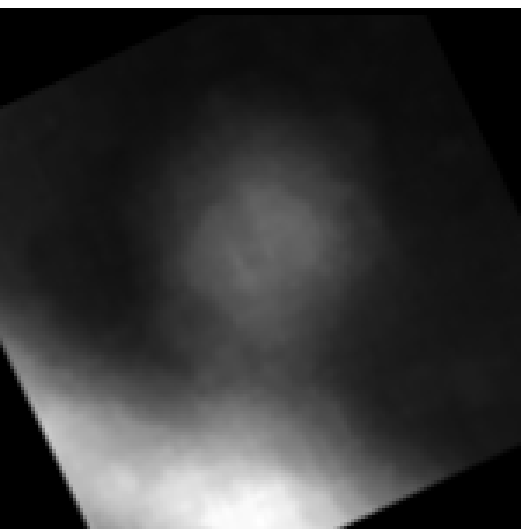} \\
\includegraphics[scale=0.2]{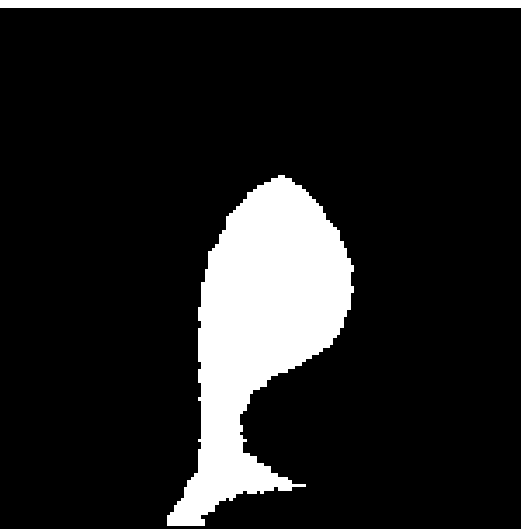}
\includegraphics[scale=0.2]{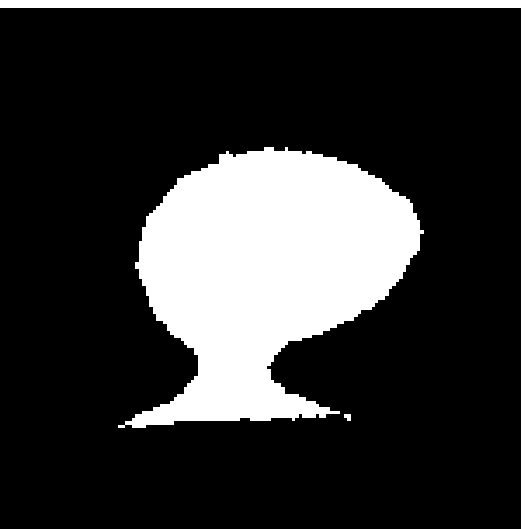}
\includegraphics[scale=0.2]{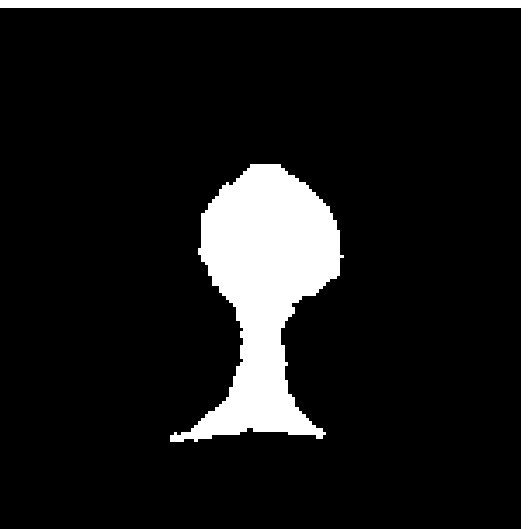}
\includegraphics[scale=0.2]{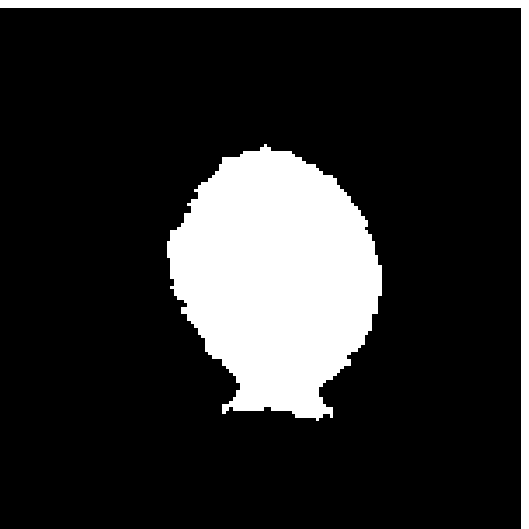}
\includegraphics[scale=0.2]{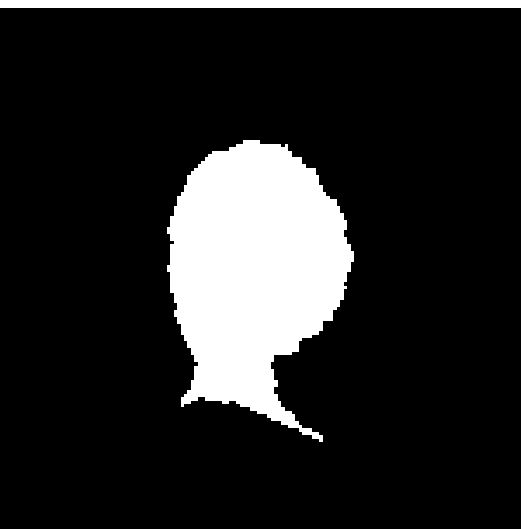} \\
\includegraphics[scale=0.2]{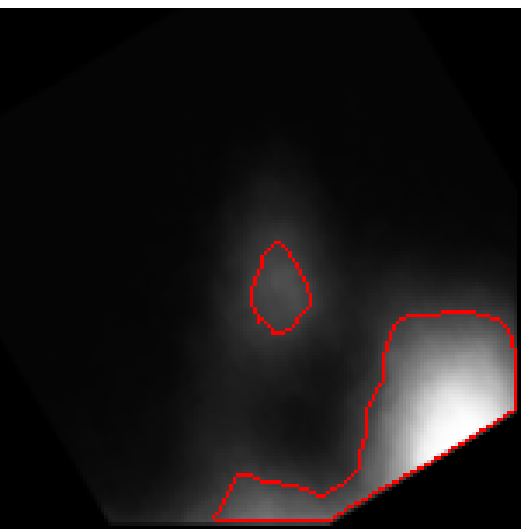}
\includegraphics[scale=0.2]{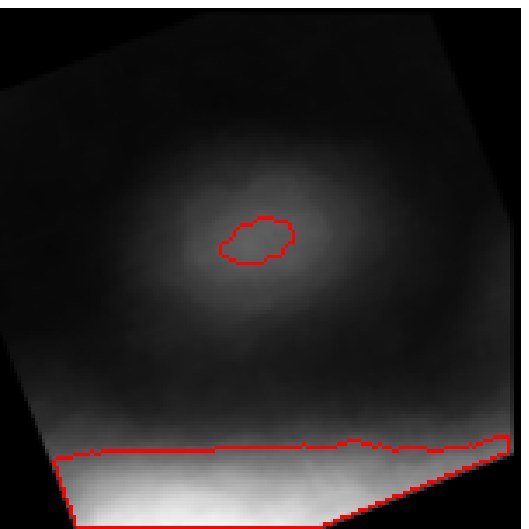}
\includegraphics[scale=0.2]{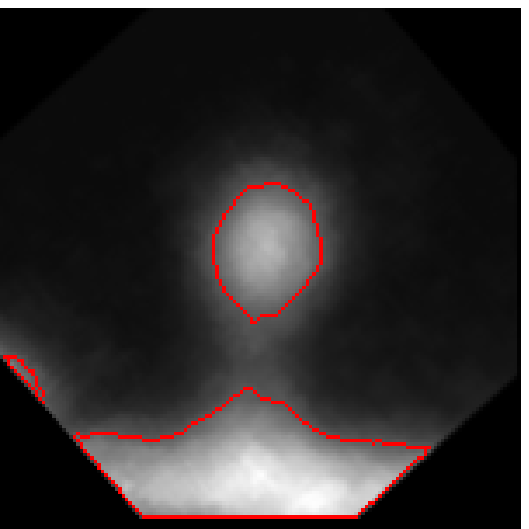}
\includegraphics[scale=0.2]{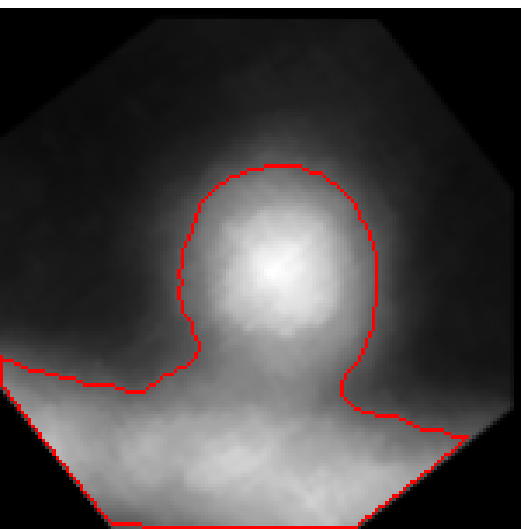}
\includegraphics[scale=0.2]{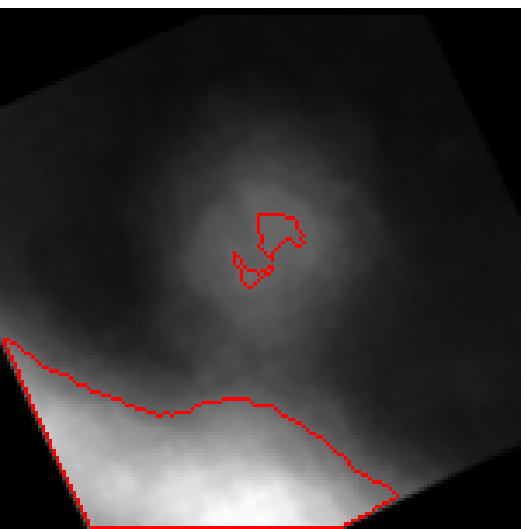} \\
\includegraphics[scale=0.2]{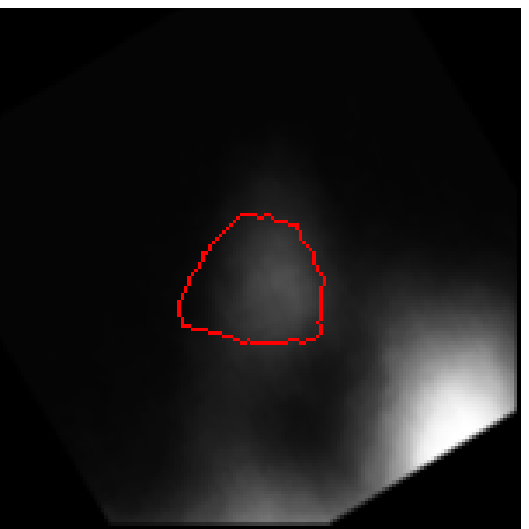}
\includegraphics[scale=0.2]{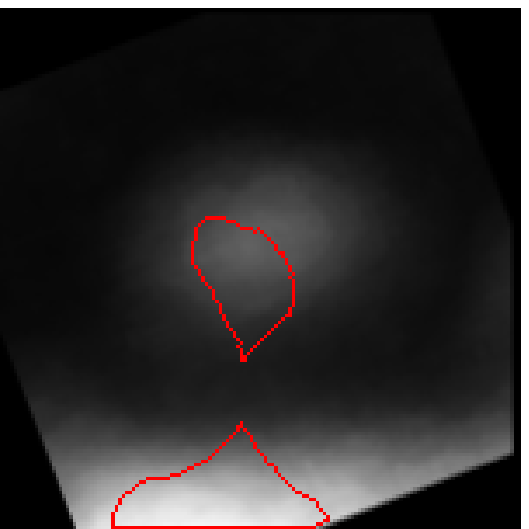}
\includegraphics[scale=0.2]{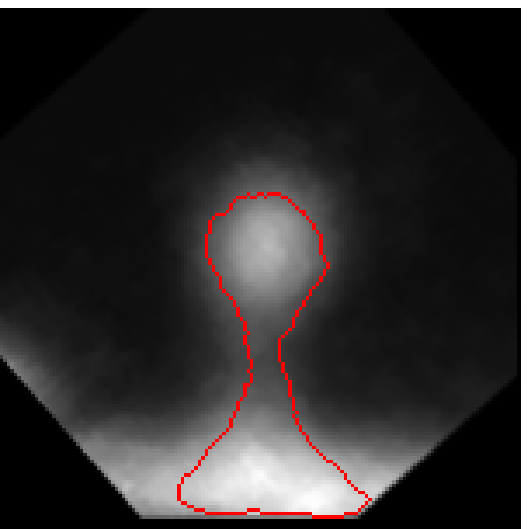}
\includegraphics[scale=0.2]{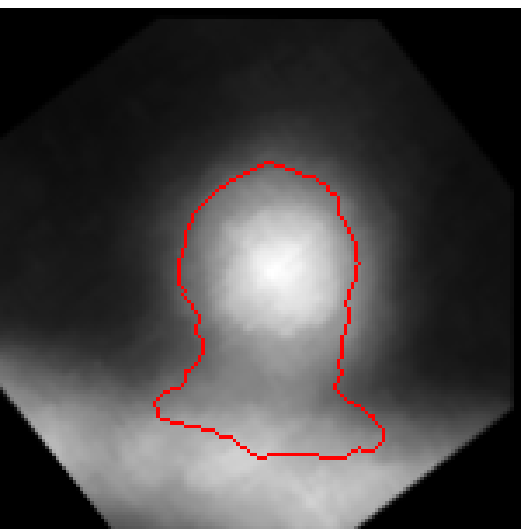}
\includegraphics[scale=0.2]{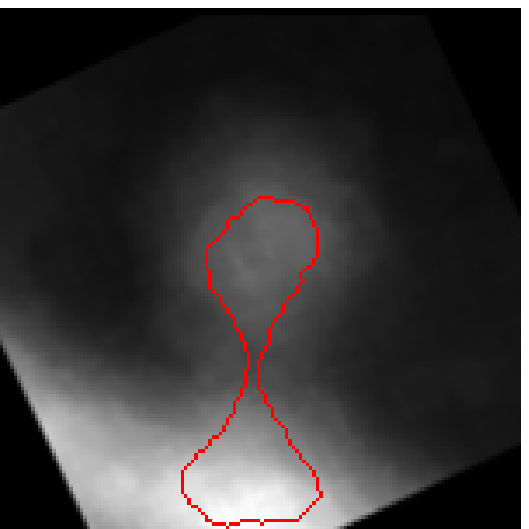} \\
\includegraphics[scale=0.2]{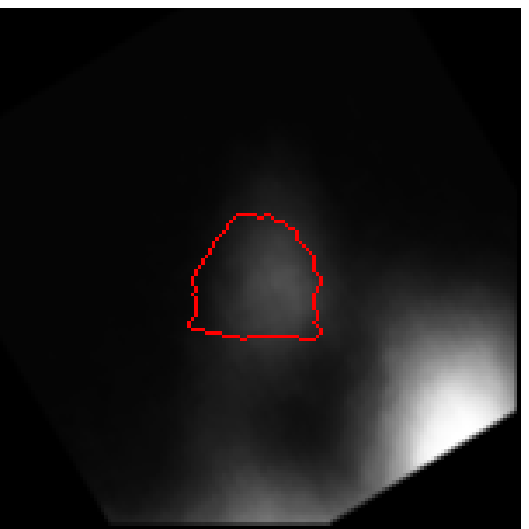}
\includegraphics[scale=0.2]{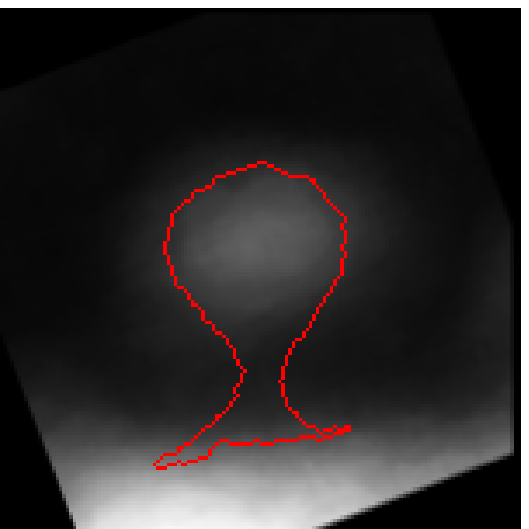}
\includegraphics[scale=0.2]{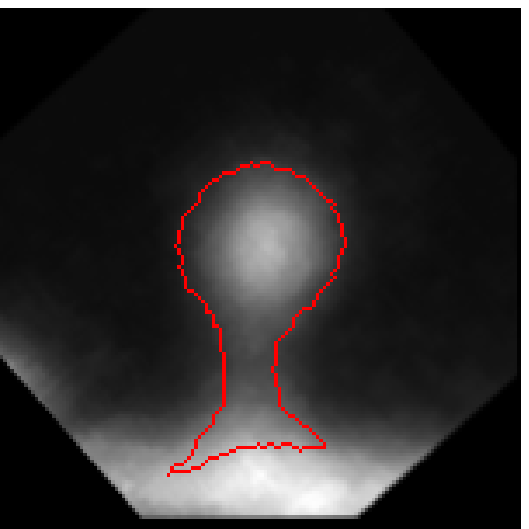}
\includegraphics[scale=0.2]{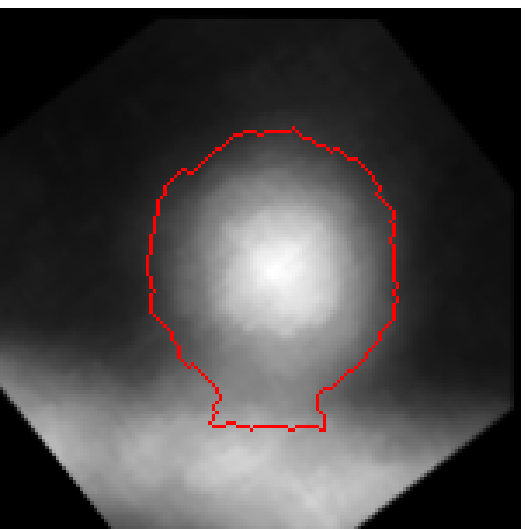}
\includegraphics[scale=0.2]{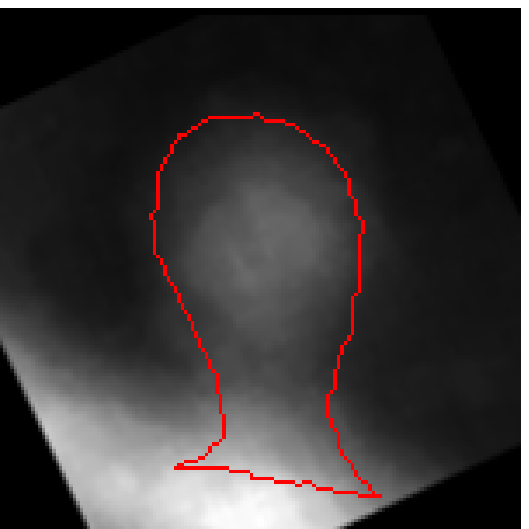} \\
\includegraphics[scale=0.2]{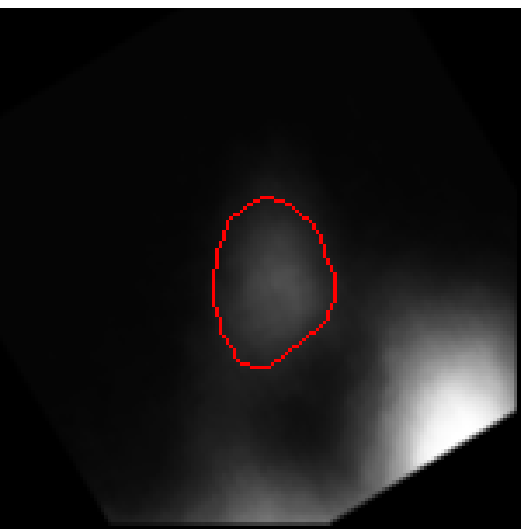}
\includegraphics[scale=0.2]{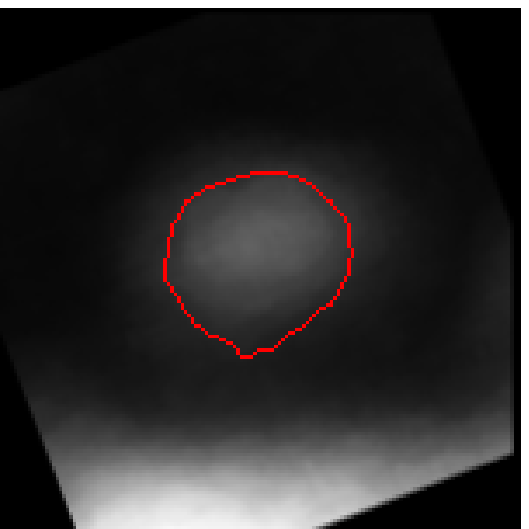}
\includegraphics[scale=0.2]{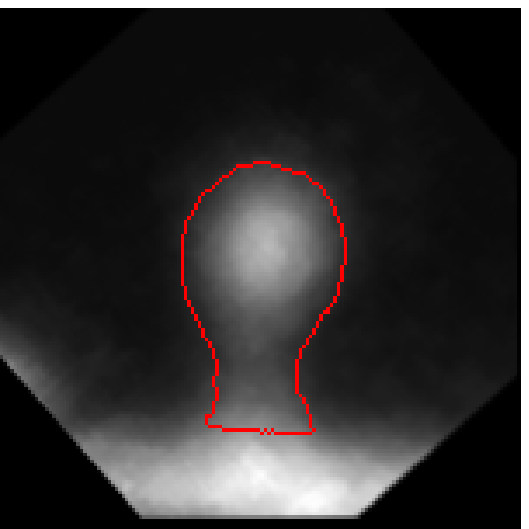}
\includegraphics[scale=0.2]{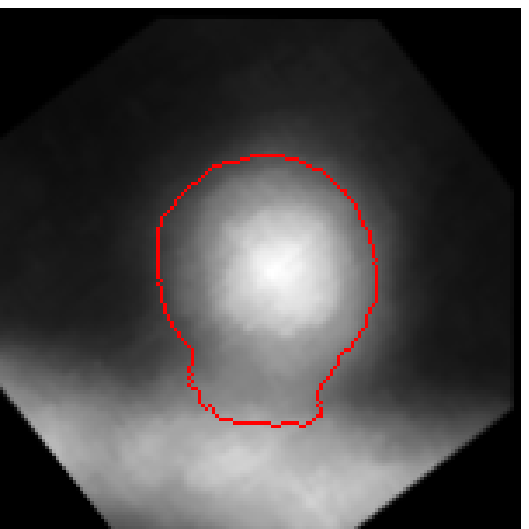}
\includegraphics[scale=0.2]{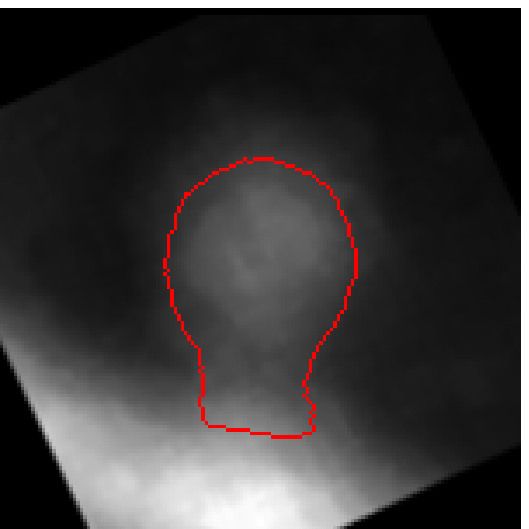} \\
\includegraphics[scale=0.2]{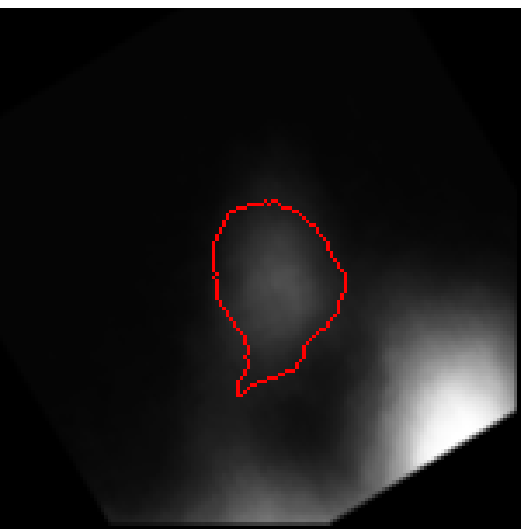}
\includegraphics[scale=0.2]{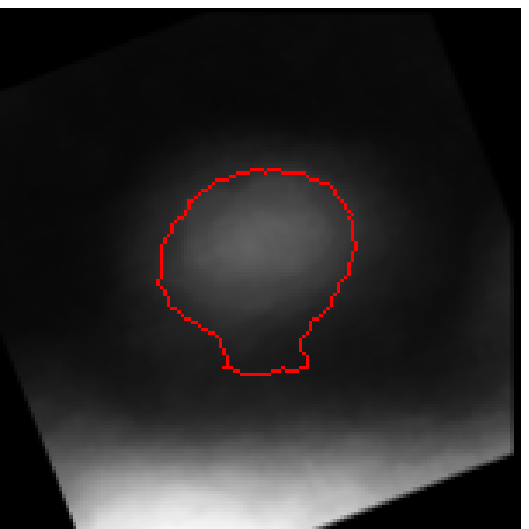}
\includegraphics[scale=0.2]{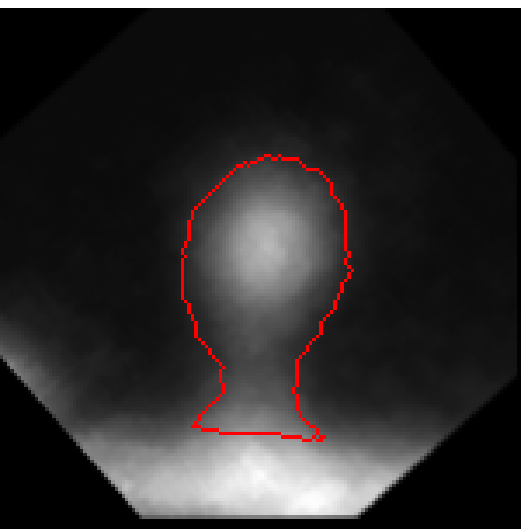}
\includegraphics[scale=0.2]{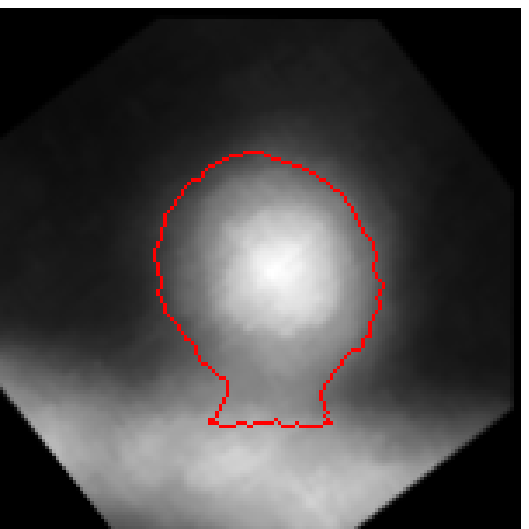}
\includegraphics[scale=0.2]{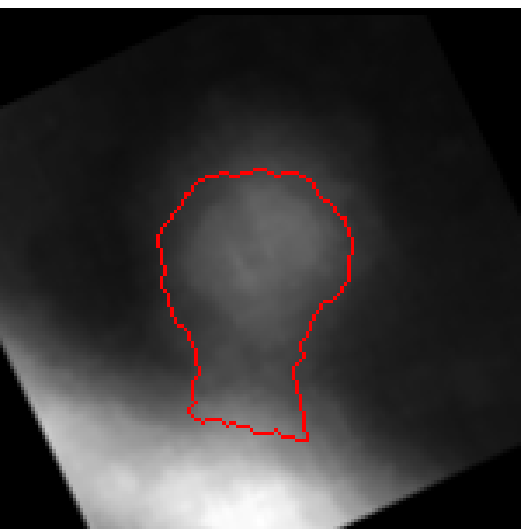} \\
\caption{Visual segmentation results on 2$D$ dendritic spine data set. First row: test image, second row: ground truth, third row: the method in \cite{chan2001active}, fourth row: the method in \cite{kim2007nonparametric}, fifth row: the method in \cite{souganli2014combining}, sixth row: segmentation with context priors and learned-intensity-based data term, seventh row: the proposed method. \label{fig:visual_results}}
\end{figure}
\section{Experimental results}
In this section, we present experimental results of the proposed approach on the dendritic spine segmentation problem. We perform experiments on both 2$D$ and 3$D$ dendritic spine data sets \cite{erdil2017nonparametric}. The dendritic spine data sets are obtained from the Neuronal Structure and Function Laboratory of the Champalimaud Neuroscience Foundation, Lisbon.

Both 2$D$ and 3$D$ dendritic spine data sets contain 30 intensity images with manual segmentations. We perform experiments on these data sets on a leave-one-out basis; one image for test and remaining 29 for training. We compare the performance of the proposed approach which uses the learned-intensity-based data term, as well as spatial context and shape priors with the approaches proposed in \cite{chan2001active}, \cite{kim2007nonparametric}, and \cite{souganli2014combining} in terms of Dice score \cite{dice1945measures}. We also obtain Dice score results for the segmentations obtained by using spatial context priors with the learned-intensity-based data term which we use as the initial step of our proposed approach as discussed before. Note that Dice score takes values between 0 and 1 where higher is better. 

The average Dice score results of 30 spines obtained from the experiments on the 2$D$ and 3$D$ dendritic spine data sets are shown in Table \ref{tab:dice_results}. We first discuss the 3$D$ results. The method in \cite{chan2001active} does not use any shape prior information; it just uses a data term that assumes the foreground and the background intensities are homogeneous. Therefore, this method produces the worst results among all methods that we use in our evaluations. The method in \cite{kim2007nonparametric} uses a nonparametric shape prior term together with the data term in \cite{chan2001active}. Although, this leads to some improvement over the results of the method in \cite{chan2001active}, the results of \cite{kim2007nonparametric} are still around $0.4319$ for the 3$D$ case in terms of Dice score. The method in \cite{souganli2014combining} uses the learned-intensity-based data term together with the nonparametric shape priors term in \cite{kim2007nonparametric}. This slightly improves the average Dice score result of \cite{kim2007nonparametric}. Among all approaches, the proposed approach which uses the learned-intensity-based data term, as well as spatial context and shape priors produces segmentations with the highest Dice score results averaged over 30 spine images. Performing segmentation using spatial context priors together with the learned intensity-based data term (the proposed approach without shape priors) produces better results than \cite{chan2001active}, \cite{kim2007nonparametric}, and \cite{souganli2014combining}, demonstrating the positive impact of the context prior term developed in this paper. Finally, we performed a statistical significance analysis comparing the Dice score results of the proposed approach and each of the competing methods using a t-test. The statistical tests demonstrate that the difference between the Dice score results of the proposed and other methods is statistically significant at the $5\%$ significance level for the 3D dendritic spine segmentation experiment.

The relative performance of the methods on the 2$D$ dendritic spine data set is very similar to the 3$D$ case in terms of Dice score as shown in Table \ref{tab:dice_results}. The proposed approach achieves the best results in terms of Dice score averaged over 30 spine images. As in the 3$D$ case, Dice score results obtained by using context priors with the learned-intensity-based data term are better than those of the existing methods. The results obtained by methods in \cite{chan2001active} and \cite{kim2007nonparametric} produce the lowest and the second lowest Dice score results, respectively. Finally, average Dice score achieved by using the method in \cite{souganli2014combining} is around 0.7347 which is lower than the methods that use context priors and higher than the other two methods. We performed statistical significance analysis using a t-test on the Dice score results of the 2$D$ segmentation experiment as well. Statistical significance analysis results demonstrate that the Dice score results of the proposed approach and the methods in \cite{chan2001active}, \cite{kim2007nonparametric}, and \cite{souganli2014combining} are statistically significant at the $5\%$ significance level. However, there is no statistically significant difference between the Dice score results of the proposed approach and the approach that uses spatial context priors and the learned-intensity-based data term. Some visual segmentation results obtained by all these approaches are shown in Figure \ref{fig:visual_results}. Note that visual segmentation results support our quantitative analysis.

\section{Conclusion}
We have proposed a segmentation approach that uses spatial context priors together with shape priors and a learned-intensity-based data term. This is a principled probabilistic approach for incorporation of information about the likely locations of particular regions in the image, within the context of a shape-prior-based segmentation framework. Assuming such statistical information about spatial context can be extracted from training data, this approach would particularly be of value in challenging segmentation problems which exhibit region-based intensity distributions with significant overlaps, making the problem harder. In this paper we have demonstrated the benefits provided by this approach in the context of microscopic neuroimage analysis, in particular on the problem of dendritic spine segmentation. However we believe the approach can also be useful in other segmentation problems, especially in biomedical or biological imaging. 

\scriptsize
\bibliographystyle{IEEEbib}
\bibliography{refs}

\begin{thebibliography}{10}

\bibitem{erdil2017nonparametric}
Ertunc Erdil, Muhammad~Usman Ghani, Lavdie Rada, Ali~Ozgur Argunsah, Devrim
  Unay, Tolga Tasdizen, and Mujdat Cetin,
\newblock ``Nonparametric joint shape and feature priors for image
  segmentation,''
\newblock {\em IEEE Transactions on Image Processing}, vol. 26, no. 11, pp.
  5312--5323, 2017.

\bibitem{cremers2006kernel}
Daniel Cremers, Stanley~J Osher, and Stefano Soatto,
\newblock ``Kernel density estimation and intrinsic alignment for shape priors
  in level set segmentation,''
\newblock {\em International Journal of Computer Vision}, vol. 69, no. 3, pp.
  335--351, 2006.

\bibitem{kim2007nonparametric}
Junmo Kim, M{\"u}jdat {\c{C}}etin, and Alan~S Willsky,
\newblock ``Nonparametric shape priors for active contour-based image
  segmentation,''
\newblock {\em Signal Processing}, vol. 87, no. 12, pp. 3021--3044, 2007.

\bibitem{kass1988snakes}
Michael Kass, Andrew Witkin, and Demetri Terzopoulos,
\newblock ``Snakes: Active contour models,''
\newblock {\em International Journal of Computer Vision}, vol. 1, no. 4, pp.
  321--331, 1988.

\bibitem{tsai2003shape}
Andy Tsai, Anthony Yezzi, William Wells, Clare Tempany, Dewey Tucker, Ayres
  Fan, W~Eric Grimson, and Alan Willsky,
\newblock ``A shape-based approach to the segmentation of medical imagery using
  level sets,''
\newblock {\em IEEE Transactions on Medical Imaging}, vol. 22, no. 2, pp.
  137--154, 2003.

\bibitem{erdil2016mcmc}
Ertunc Erdil, Sinan Yildirim, Mujdat Cetin, and Tolga Tasdizen,
\newblock ``Mcmc shape sampling for image segmentation with nonparametric shape
  priors,''
\newblock in {\em Proceedings of the IEEE Conference on Computer Vision and
  Pattern Recognition}, 2016, pp. 411--419.

\bibitem{souganli2014combining}
Abdurrahim So{\u{g}}anl{\i}, Mustafa~G{\"o}khan Uzunba{\c{s}}, and M{\"u}jdat
  {\c{C}}etin,
\newblock ``Combining learning-based intensity distributions with nonparametric
  shape priors for image segmentation,''
\newblock {\em Signal, Image and Video Processing}, vol. 8, no. 4, pp.
  789--798, 2014.

\bibitem{fedkiw2002level}
Stanley Osher~Ronald Fedkiw and Stanley Osher,
\newblock ``Level set methods and dynamic implicit surfaces,''
\newblock {\em Surfaces}, vol. 44, pp. 77, 2002.

\bibitem{chan2001active}
Tony~F Chan and Luminita~A Vese,
\newblock ``Active contours without edges,''
\newblock {\em IEEE Transactions on Image Processing}, vol. 10, no. 2, pp.
  266--277, 2001.

\bibitem{dice1945measures}
Lee~R Dice,
\newblock ``Measures of the amount of ecologic association between species,''
\newblock {\em Ecology}, vol. 26, no. 3, pp. 297--302, 1945.

\end{thebibliography}

\end{document}